\documentclass{JHEP3}
\pdfoutput=1 
\usepackage{amsmath}
\usepackage{cite}
\usepackage{epsfig,multicol}

\def\d{\mathrm{d}}

\title{Extracting the ${\boldsymbol{\rho}}$ meson wavefunction from HERA data}
\author{J.R. Forshaw \\
School of Physics \& Astronomy, University of Manchester, \\
Oxford Road, Manchester M13 9PL, U.K.\\
\email{jeff.forshaw@manchester.ac.uk} }
\author{R. Sandapen \\
D\'epartement de Physique et d'Astronomie, Universit\'e de Moncton, \\
Moncton, N-B. E1A 3E9, Canada.\\
\email{ruben.sandapen@umoncton.ca} }
\preprint{MAN/HEP/2010/8}
\abstract{
We extract the light-cone wavefunctions of the $\rho$ meson using the
HERA data on diffractive $\rho$ photoproduction. We find good
agreement with predictions for the distribution amplitude based on QCD
sum rules and from the lattice. We also find that the
data prefer a transverse wavefunction with enhanced end-point contributions. 
}
\keywords{QCD, diffraction, light-cone wavefunction}

\begin{document}
\section{Introduction}

Diffractive $\rho$ meson photoproduction\footnote{We use
  `photoproduction' to refer to production using both real and virtual
  photons.}  can be described within the dipole model
\cite{Nikolaev:1990ja,Nikolaev:1991et,Mueller:1993rr,Mueller:1994jq}
according to the formula
\begin{equation}
\left.\Im \mbox{m} \, \mathcal{A}_\lambda(s,t)\right|_{t=0} = s \sum_{h,
\bar{h}}
\int \d^2 {\mathbf r} \; \d z \; \Psi^{\gamma,\lambda}_{h, \bar{h}}(r,z)
\hat{\sigma}(s,r) \Psi^{\rho,\lambda}_{h, \bar{h}}(r,z)^*
\label{amplitude}
\end{equation}
for the imaginary part of the forward elastic scattering amplitude,
where $\Psi^{\gamma,\lambda}_{h, \bar{h}}(r,z)$ and
$\Psi^{\rho,\lambda}_{h, \bar{h}}(r,z)$  are the light-cone
wavefunctions of the photon and vector meson, and $\hat{\sigma}(s,r)$
is the dipole cross section. The light-cone wavefunctions represent
the probability amplitudes for the photon or $\rho$ meson to fluctuate
into a $q\bar{q}$ pair of transverse size $r$ in which the quark
carries a fraction $z$ of the meson's light-cone momentum. The sum is
over quark/antiquark helicities ($h$ and $\bar{h}$) and $\lambda=L~\mbox{or}~T$
labels the polarization of the photon and meson.

The photoproduction cross section is,
after integrating over $t$,
\begin{equation}
 \sigma_\lambda(s)  = \frac{1}{B}
\frac
{1}{16\pi} (\Im\mathrm{m} \mathcal{A}_\lambda(s,0))^2 \; (1 + \beta_\lambda^2)~,
\label{gammap-xsec}
\end{equation}
where $\beta_\lambda$ is the ratio of real to imaginary parts of the
amplitude and $B$ is the diffractive slope\footnote{We assume $\d
  \sigma /\d t \propto \exp(Bt)$.}. We shall assume that
\begin{equation}
B=N\left(
  14.0 \left(\frac{1~\mathrm{GeV}^2}{Q^2 + M_{\rho}^2}\right)^{0.2}+1\right)
\label{Bslope}
\end{equation}
with $N=0.55$ GeV$^{-2}$, which is in accord with the ZEUS data
\cite{Chekanov:2007zr}. The H1 data \cite{Collaboration:2009xp}
prefer a somewhat larger value of $B$, but with a larger
uncertainty. We compute
$\beta_\lambda$ according to
\begin{equation}
 \beta_\lambda=\tan \left(\frac{\pi}{2} \alpha_\lambda \right) 
\end{equation}
with
\begin{equation}
 \alpha_\lambda=\frac{\partial \ln |\Im \mathrm{m}\,
\mathcal{A}_\lambda|}{\partial \ln \left(1/x\right)}
\label{logderivative}
\end{equation}
where $\Im \mathrm{m}\, \mathcal{A}_\lambda$ is given by
Eq.~(\ref{amplitude}) and $x = (Q^2 + 4m_f^2)/(Q^2+ W^2)$ ($m_f$ is
defined in the next section). The total cross section that is measured
experimentally is given by
\begin{equation}
 \sigma=\sigma_T + \epsilon \sigma_L 
\end{equation}
where $\epsilon=0.98$. 

Our goal in this paper is to use the current HERA data on $\rho$-meson
photoproduction \cite{Chekanov:2007zr, Collaboration:2009xp} to extract the
meson's light-cone wavefunction. To do that we must specify the dipole cross
section, which we do by assuming the FS2004 saturation model that was
extracted from the HERA deep inelastic scattering data in
Ref.~\cite{Forshaw:2004vv}. 

\section{Light-cone wavefunctions}
The photon's light-cone wavefunctions are
\cite{Lepage:1980fj,Dosch:1996ss,Kulzinger:1998hw}:
\begin{equation}
\Psi^{\gamma,L}_{h,\bar{h}}(r,z) = \sqrt{\frac{N_{c}}{4\pi}}\delta_{h,-\bar{h}}e
e_{f}2 z(1-z) Q \frac{K_{0}(\epsilon r)}{2\pi}\;,
\label{photonwfL}
\end{equation}
and 
\begin{equation}
\Psi^{\gamma,T}_{h,\bar{h}}(r,z) = \pm \sqrt{\frac{N_{c}}{2\pi}} ee_{f}
 \big[i e^{ \pm i\theta_{r}} (z \delta_{h\pm,\bar{h}\mp} - 
(1-z) \delta_{h\mp,\bar{h}\pm}) \partial_{r}  
+  m_{f} \delta_{h\pm,\bar{h}\pm} \big]\frac{K_{0}(\epsilon r)}{2\pi}\;,
\label{photonwfT}
\end{equation}
where 
\begin{equation}
\epsilon^{2} = z(1-z)Q^{2} + m_{f}^{2}~.
\end{equation}
These wavefunctions are derived from perturbative QED and depend upon
a phenomenological light-quark mass, $m_f$. We are compelled to take
$m_f=0.14$ GeV, as determined by the fit in
Ref.~\cite{Forshaw:2004vv}. We take $e_f = 1/\sqrt{2}$, as appropriate
for $\rho$ meson production.

The meson's light-cone wavefunctions can be written in terms of the scalar
wavefunctions $\phi_{L,T}(r,z)$:
\begin{equation}
\Psi^{\rho,L}_{h,\bar{h}}(r,z) = \sqrt{\frac{N_{c}}{4\pi}}
\delta_{h,-\bar{h}}\frac{1}{M_{\rho}z(1-z)} 
[z(1-z)M^{2}_{\rho} +  m_{f}^{2} -  \nabla_{r}^{2}] \phi_L(r,z) 
\label{nnpz_L}
\end{equation}
where $\nabla_r^2 \equiv \frac{1}{r} \partial_r + \partial^2_r$ and 
\begin{equation}
\Psi^{\rho,T}_{h,\bar{h}}(r, z) = \pm
\sqrt{\frac{N_{c}}{4\pi}}\frac{\sqrt{2}}{z(1-z)}[i e^{\pm i\theta_{r}} 
( z \delta_{h\pm,\bar{h}\mp} - (1-z) \delta_{h\mp,\bar{h}\pm}) 
\partial_{r}+ m_{f}\delta_{h\pm,\bar{h}\pm}] \phi_T(r, z).
\label{nnpz_T}
\end{equation}
These wavefunctions are subject to two important constraints. The first is the
normalisation condition, which embodies our assumption that the $\rho$ meson
consists solely of a $q\bar{q}$ pair:

\begin{equation}
\sum_{h,\bar{h}}\int \d^{2}{\mathbf{r}} \, \d z  \,
|\Psi^{\rho,\lambda}_{h,\bar{h}}(r, z)|^{2} = 1 ~.
\label{normalisation}
\end{equation}
The second constraint arises from the measured value of $f_{\rho}$, the meson
decay constant for the longitudinally polarised meson, i.e.
\begin{equation}
f_\rho M_\rho = \frac{N_c}{\pi} {e}_f \int_0^1 \frac{\d z}{z(1-z)} 
\left.[z(1-z)M^{2}_{\rho} + m_{f}^{2}-\nabla_{r}^{2}] \phi_L(r,z)
\right|_{r=0}~. 
\label{longdecay}
\end{equation}
The decay constant is deduced from the experimentally measured
electronic decay
width via the relation \cite{Kulzinger:1998hw}: 
\begin{equation}
 \Gamma_{\rho \rightarrow e^{+}e^{-}}=\frac{4 \pi \alpha_{\mathrm{em}}^2
f_\rho^2}{3 M_{\rho}}~,
\end{equation}
where $\Gamma_{\rho \rightarrow e^{+}e^{-}}=7.04 \pm 0.06~\mathrm{keV}$
\cite{Nakamura:2010zzi}.

\section{Fitting the HERA data}
We must specify the form of the scalar wavefunctions $\phi_{\lambda}$. In
Ref.~\cite{Forshaw:2003ki}, a `Boosted Gaussian' (BG) wavefunction of the form
\begin{eqnarray}
\phi^{{\mathrm{BG}}}_\lambda(r,z) &=&
\mathcal{N}_\lambda \;  4[z(1-z)]^{b_{\lambda}} \sqrt{2\pi R_{\lambda}^{2}} \;
\exp \left(\frac{m_f^{2}R_{\lambda}^{2}}{2}\right)
\exp \left(-\frac{m_f^{2}R_{\lambda}^{2}}{8[z(1-z)]^{b_{\lambda}}}\right) \\
\nonumber
& &\times \exp \left(-\frac{2[z(1-z)]^{b_\lambda}
r^{2}}{R_{\lambda}^{2}}\right) 
\label{boosted-gaussian} 
\end{eqnarray}
was used.
This wavefunction is a simplified version of that proposed originally
by Nemchik, Nikolaev, Predazzi and Zakharov \cite{Nemchik:1996cw}.

\begin{figure}
\centering
\includegraphics*[width=6cm]{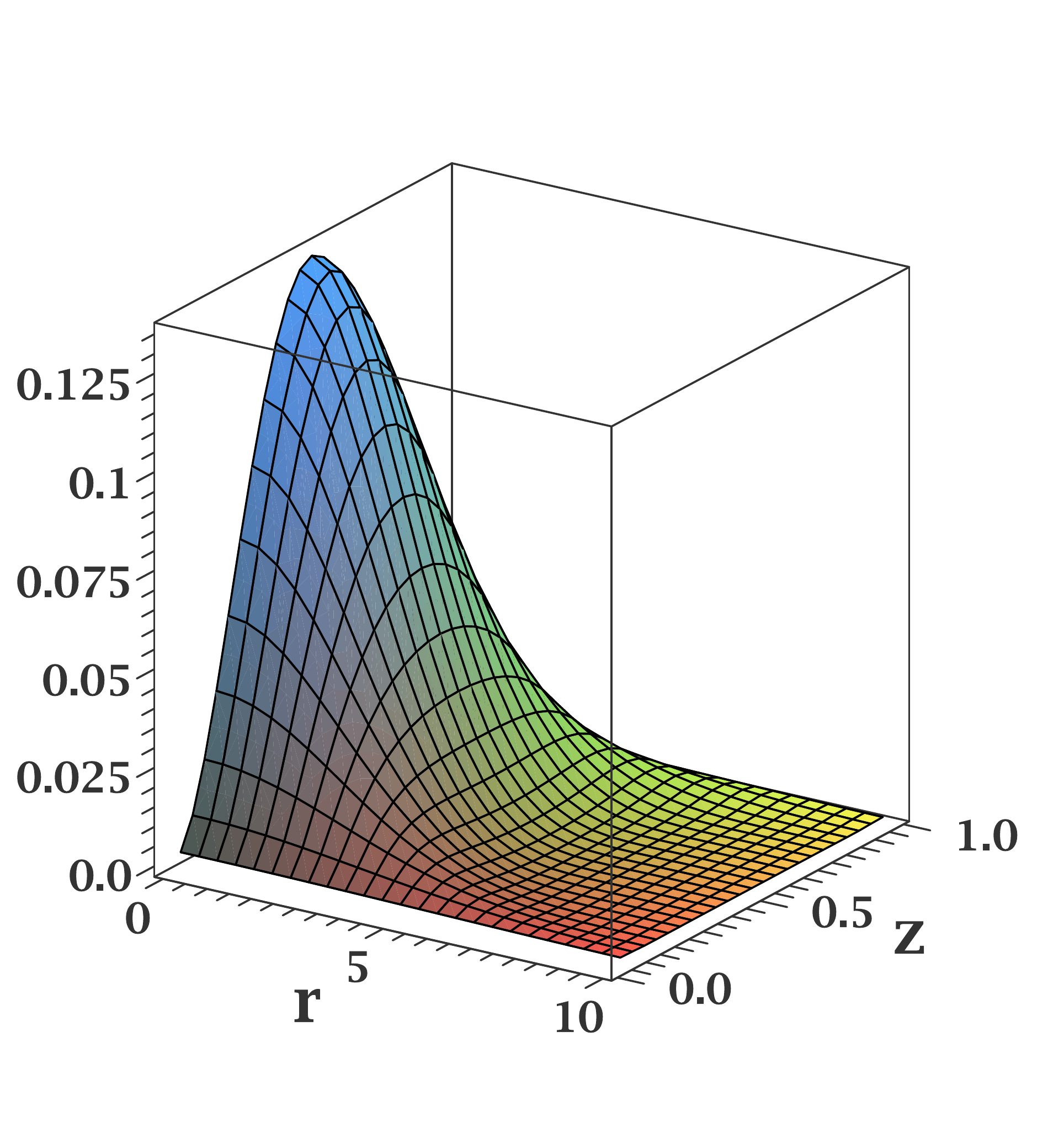}
\hspace{1cm}
\includegraphics*[width=6cm]{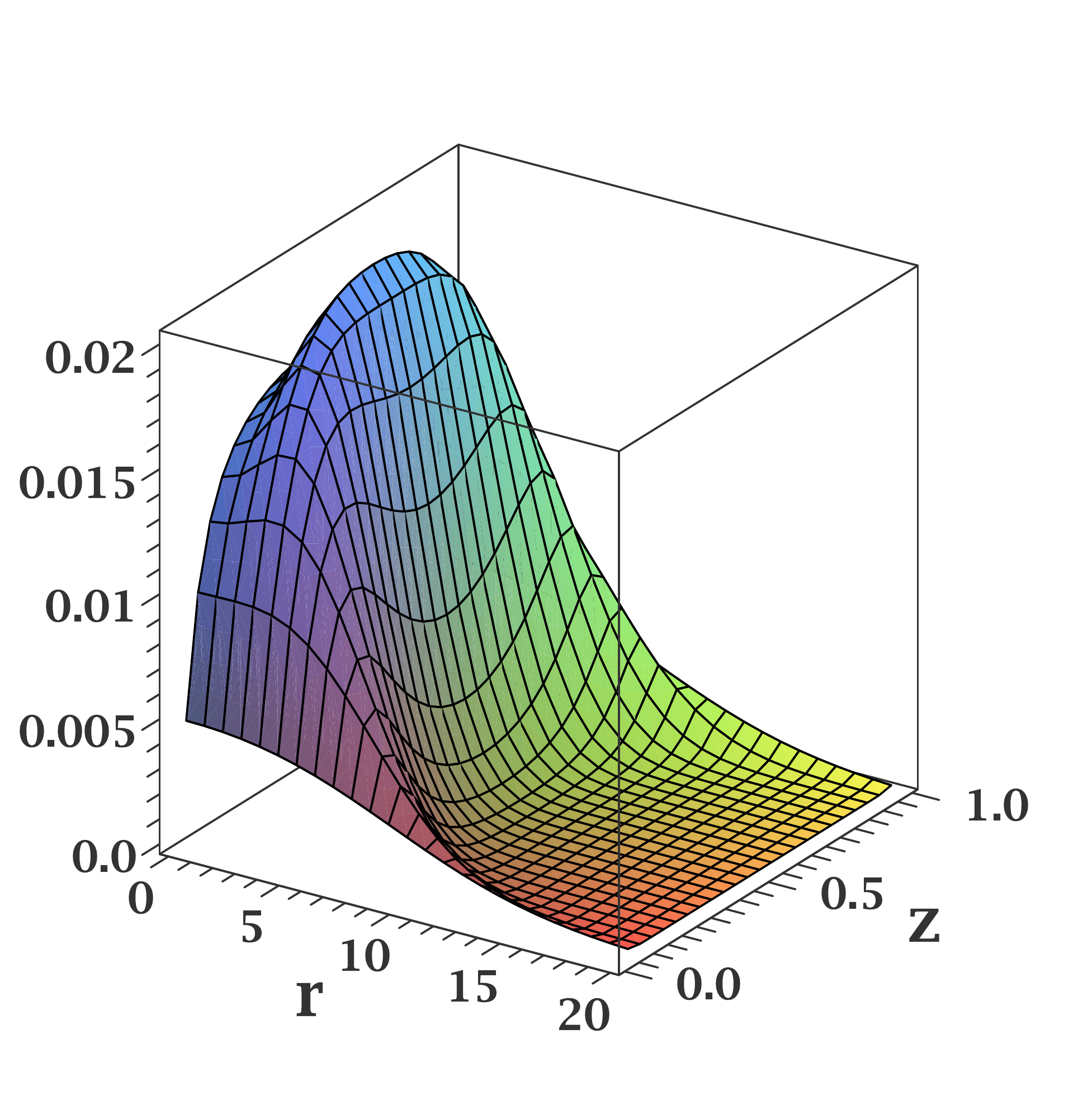}  
\caption{The modulus squared of the longitudinal (left) and transverse (right)
`Boosted Gaussian' light-cone wavefunctions from Ref.~\cite{Forshaw:2003ki}.}
\label{fig:BG_LCWF}
\end{figure}

In Ref.~\cite{Forshaw:2003ki}, it was assumed that $b_\lambda=1$ and
that $R_{L}=R_{T}=R$. The leptonic decay width constraint and the
normalization conditions fix $R$ and $\mathcal{N}_{\lambda}$
(i.e. $R^2=12.3$~GeV$^{-2}$), leaving no free parameters. Predictions
can then be made for the $\rho$-meson photoproduction
cross section. This procedure leads to reasonable agreement with the
old HERA data on the light vector mesons \cite{Forshaw:2003ki} and
also for the heavier $J/\Psi$ \cite{Forshaw:2006np}. However, it is
not able to accommodate the most recent HERA data on $\rho$
production. Comparison with the HERA data leads to a $\chi^2/$data
point of $234/75$. For comparison with our later results, the longitudinal and
transverse
BG light-cone wavefunctions are shown in Figure~\ref{fig:BG_LCWF}. 

\begin{figure}
 \centering
\includegraphics*[width=15.cm]{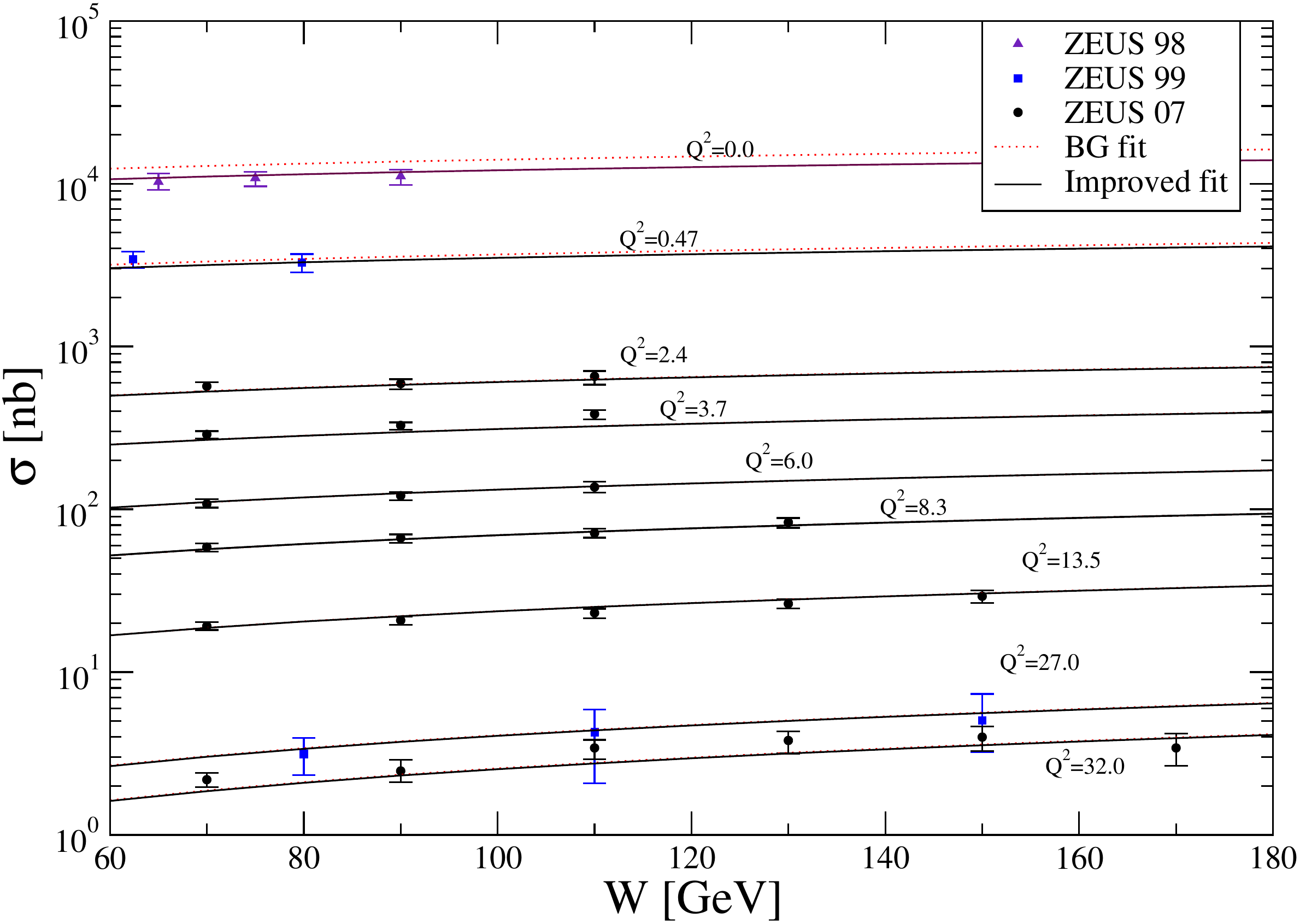}
\includegraphics*[width=15.cm]{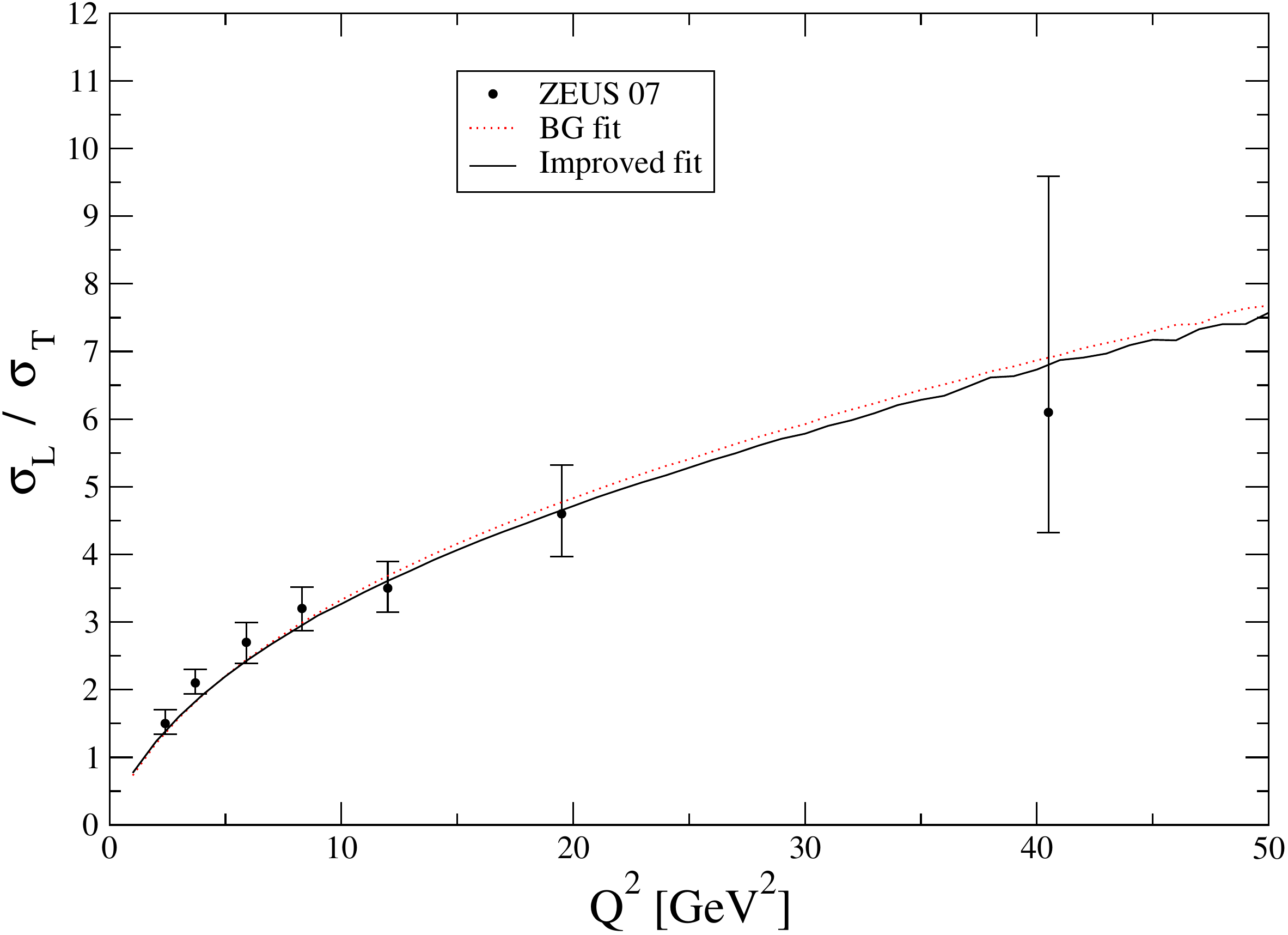}  
 \caption{Comparison to the ZEUS data. The $\sigma_L/\sigma_T$ data
   are at $W = 90$~GeV. }
\label{fig:ZEUSfits}
\end{figure}

\begin{figure}
 \centering
\includegraphics*[width=15.cm]{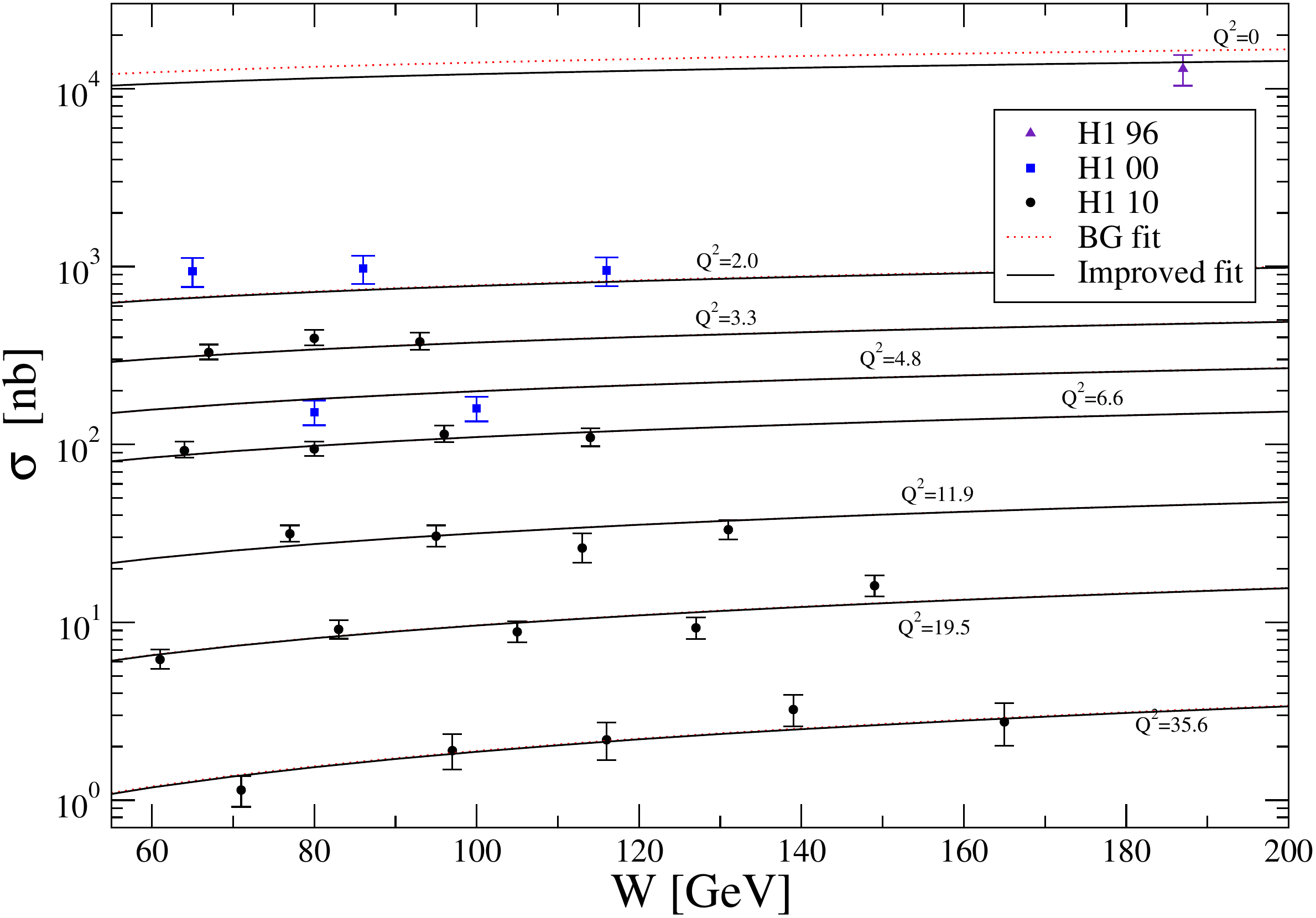}
\includegraphics*[width=15.cm]{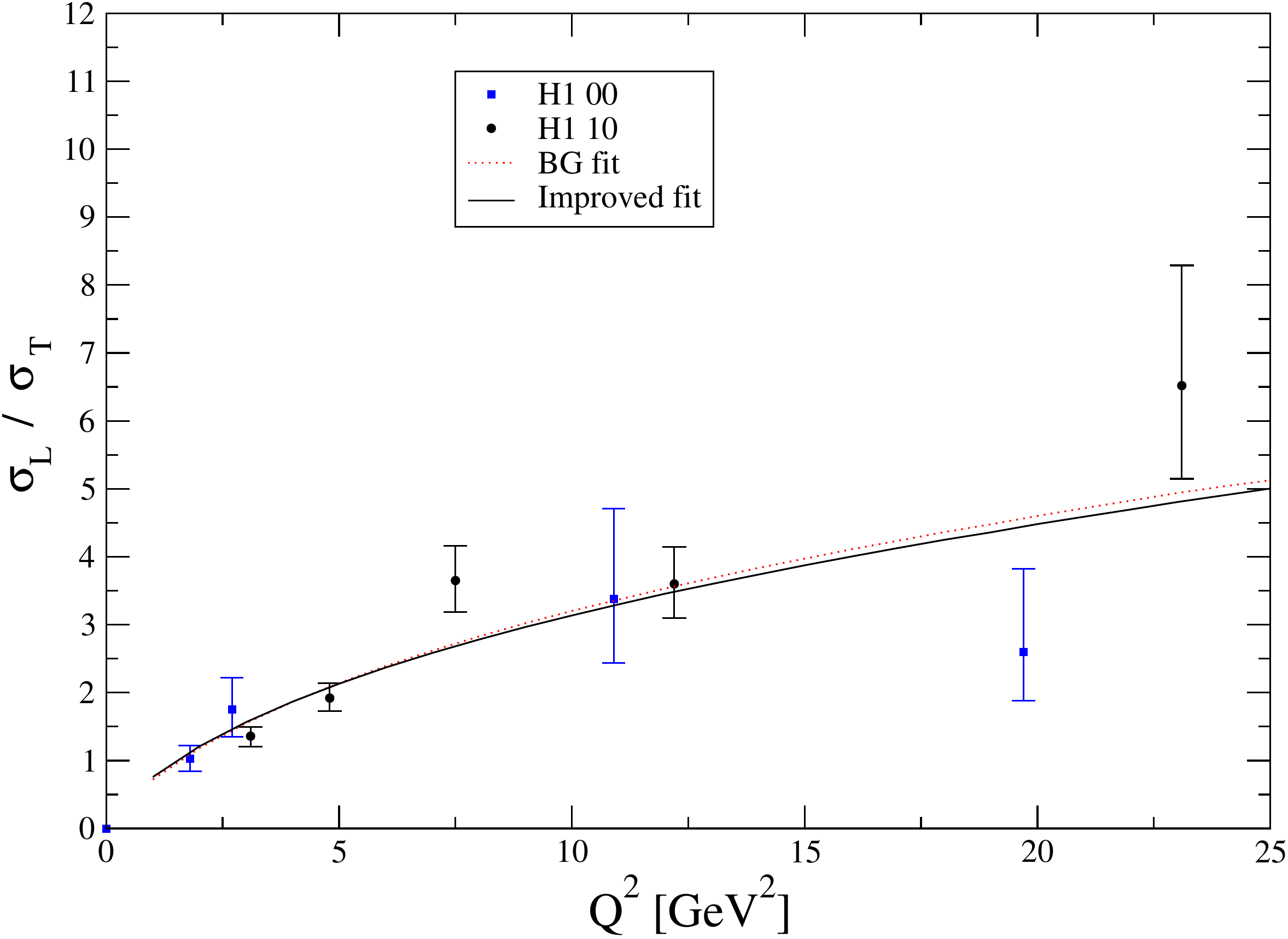}  
 \caption{Comparison to the H1 data. The $\sigma_L/\sigma_T$ data
   are at $W = 75$~GeV. }
\label{fig:H1fits}
\end{figure}

The poor agreement with data is considerably improved by allowing
$b_\lambda$ and $R_\lambda$ to vary freely.\footnote{Allowing only $R_\lambda$
to vary does not much improve the fit.}  Specifically, we fit to the
photoproduction total cross section,\footnote{We rescale the H1 and ZEUS data
  by $0.95$, which is consistent with the experimental
  uncertainty in the overall normalisation of the data
  \cite{Chekanov:2007zr,Collaboration:2009xp}.} the
longitudinal-to-transverse cross section ratio and to the electronic
decay constant datum using Minuit \cite{James:1975dr}. To ensure that we fit in the diffractive region,
we  exclude those data points with $W \le 60$~GeV. We fit to the most
recent data from H1 \cite{Collaboration:2009xp} and ZEUS
\cite{Chekanov:2007zr}. In addition, we include the earlier HERA data
\cite{Adloff:1999kg,Aid:1996bs,Breitweg:1998nh,Breitweg:1997ed} when they 
are in a kinematic region not covered by the latest data. This selected data set
comprises $39$ data points from ZEUS ($32$ points for the total
cross section and $7$ points for the longitudinal-to-transverse cross
section ratio) and $36$ data points from H1 ($27$ points for the total
cross section and $9$ points for the longitudinal-to-transverse cross
section ratio). To these, we add one data point for the decay
constant of the longitudinally polarised $\rho$ meson, giving a total of
$76$ data points. 
The result is a $\chi^2/$degree of freedom equal to  $82/72$ and the
corresponding best-fit parameters are listed in Table
\ref{tab:BGFitparams}. The results of this fit can be seen as the
dotted lines in Figures
\ref{fig:ZEUSfits} and \ref{fig:H1fits}, where the poor agreement at $Q^2 =
0$ is to be noted. 

\begin{table}[h]
\begin{center}
\textbf{Boosted Gaussian fit}
\[
\begin{array} 
[c]{|c|c|}\hline
\mbox{Free parameter} & \mbox{Fitted value}  \\ \hline
R_{L}^2 & 27.33~\mbox{GeV}^{-2}\\ \hline
R_{T}^2 & 30.87~\mbox{GeV}^{-2} \\ \hline
b_{L} & 0.5545\\ \hline
b_{T} & 0.6792\\ \hline
\end{array}
\]
\end{center}
\caption {The parameters corresponding to the BG fit of Eq.~(3.1). The
  $\chi^2/$degree of freedom is equal to  $82/72$.}
\label{tab:BGFitparams}
\end{table}

We can further improve the quality of fit (especially at $Q^2=0$) by allowing
for
additional end-point enhancement in the meson wavefunctions, i.e. using a scalar
wavefunction of the form
\begin{equation}
\phi_\lambda (r,z)= \phi^{{\mathrm{BG}}}_\lambda(r,z) \times [1+ c_{\lambda}
\xi^2 + d_{\lambda} \xi^4]
\label{EG} 
\end{equation}
where $\xi=2z-1$. We find a preference for enhancement only
in the wavefunction for transversely polarized mesons, i.e.
the data prefer $c_L=d_L=0$, which is not surprising since $\sigma_L$ vanishes
at $Q^2=0$. The new fit lowers the $\chi^2/$degree of freedom to
$68/70$. The resulting 
fitted parameters are listed in Table \ref{tab:EGFitparams} and the
corresponding cross section predictions appear as the solid lines in
Figures~\ref{fig:ZEUSfits} and \ref{fig:H1fits}.

\begin{table}[h]
\begin{center}
\textbf{Best fit}
\[
\begin{array} 
[c]{|c|c|}\hline
\mbox{Free parameter} & \mbox{Fitted value}  \\ \hline
R_{L}^2 & 26.76~\mbox{GeV}^{-2} \\ \hline
R_{T}^2 & 27.52~\mbox{GeV}^{-2} \\ \hline
b_{L} & 0.5665\\ \hline
b_{T} & 0.7468\\ \hline
c_{T} & 0.3317\\ \hline
d_{T} & 1.310\\ \hline
\end{array}
\]
\end{center}
\caption {The parameters corresponding to the BG fit with additional end-point
  enhancement in the transverse wavefunction, i.e. Eq.~(3.2). The
  $\chi^2/$degree of freedom is equal to $68/70$.}
\label{tab:EGFitparams}
\end{table}

\begin{figure}
\centering
\includegraphics*[width=7.5cm]{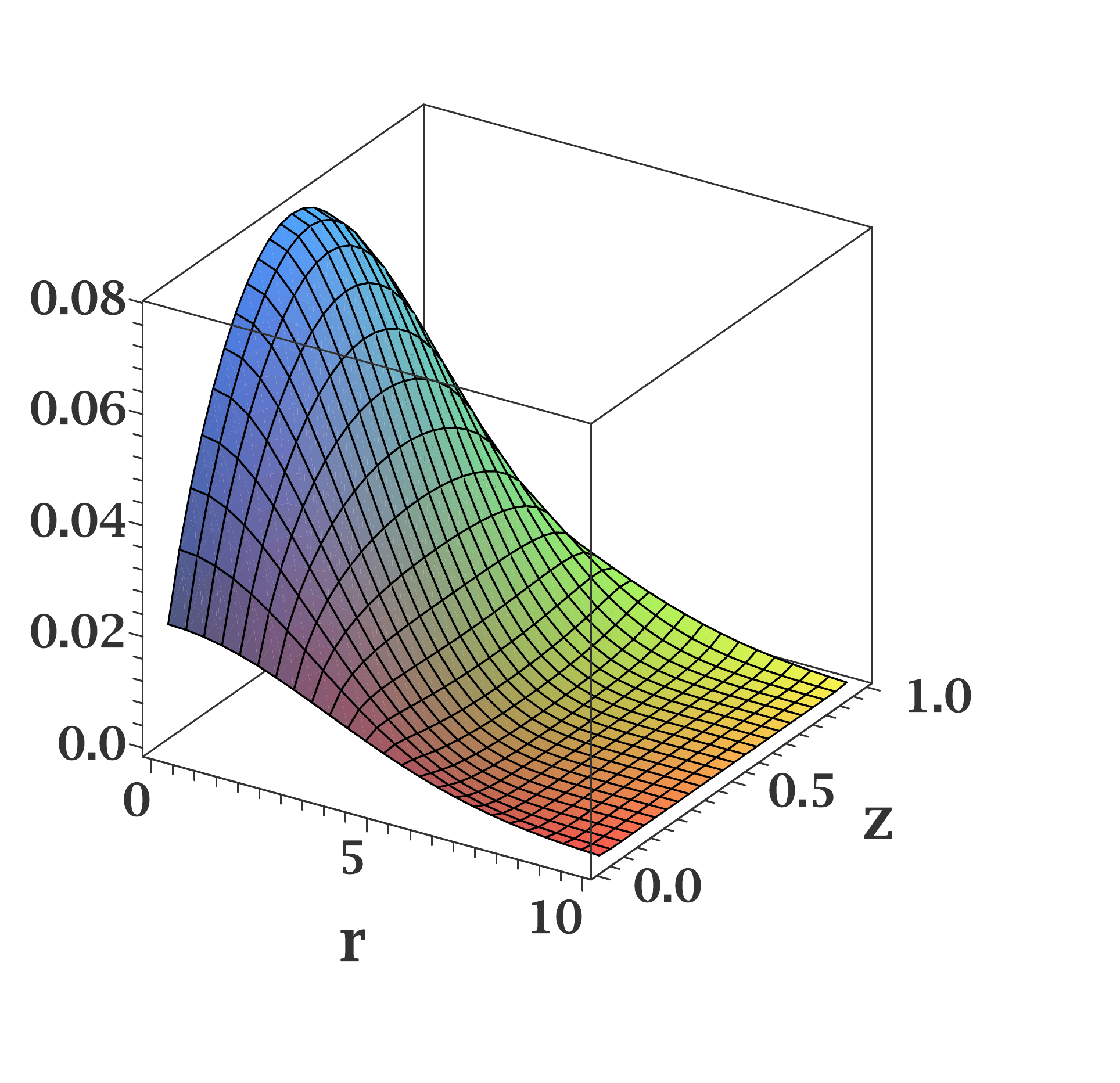}
\includegraphics*[width=7.5cm]{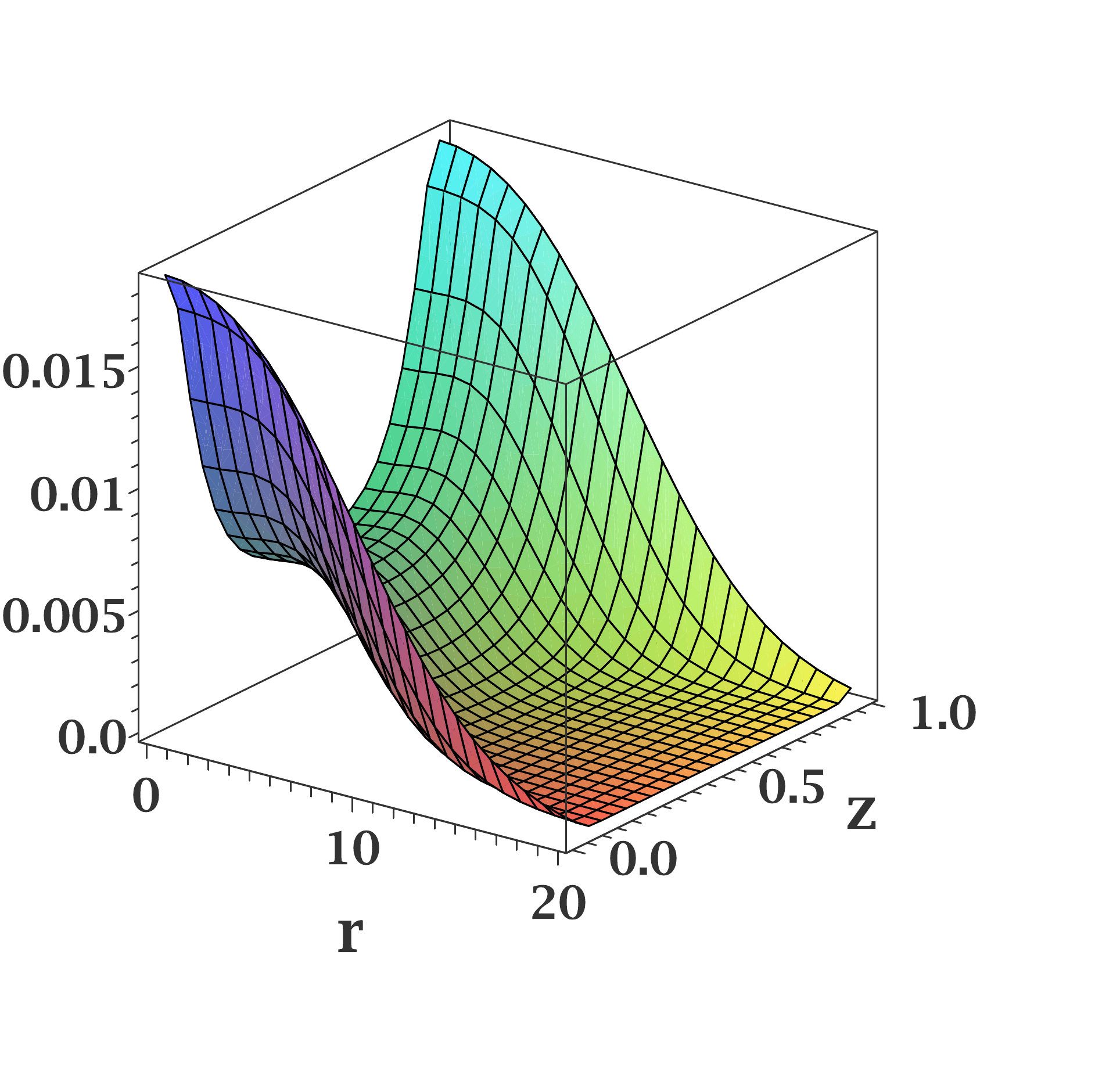}  
\caption{The longitudinal (left) and transverse (right) light-cone
  wavefunctions squared corresponding to the BG fit with additional end-point
  enhancement in the transverse wavefunction.}
\label{fig:LCWF_FitZEUS}
\end{figure}

\begin{figure}
\centering
\includegraphics*[width=7.5cm]{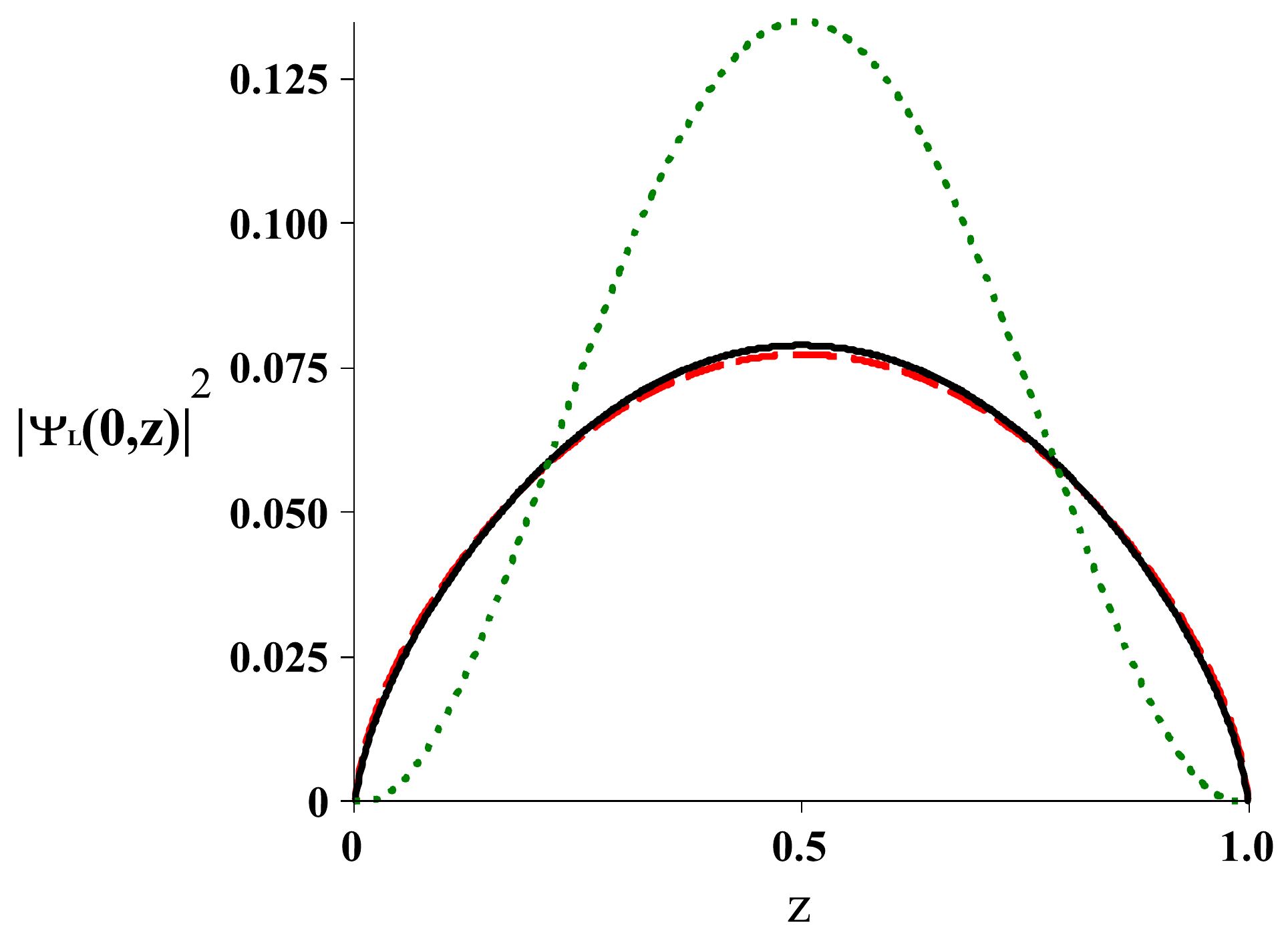}
\includegraphics*[width=7.5cm]{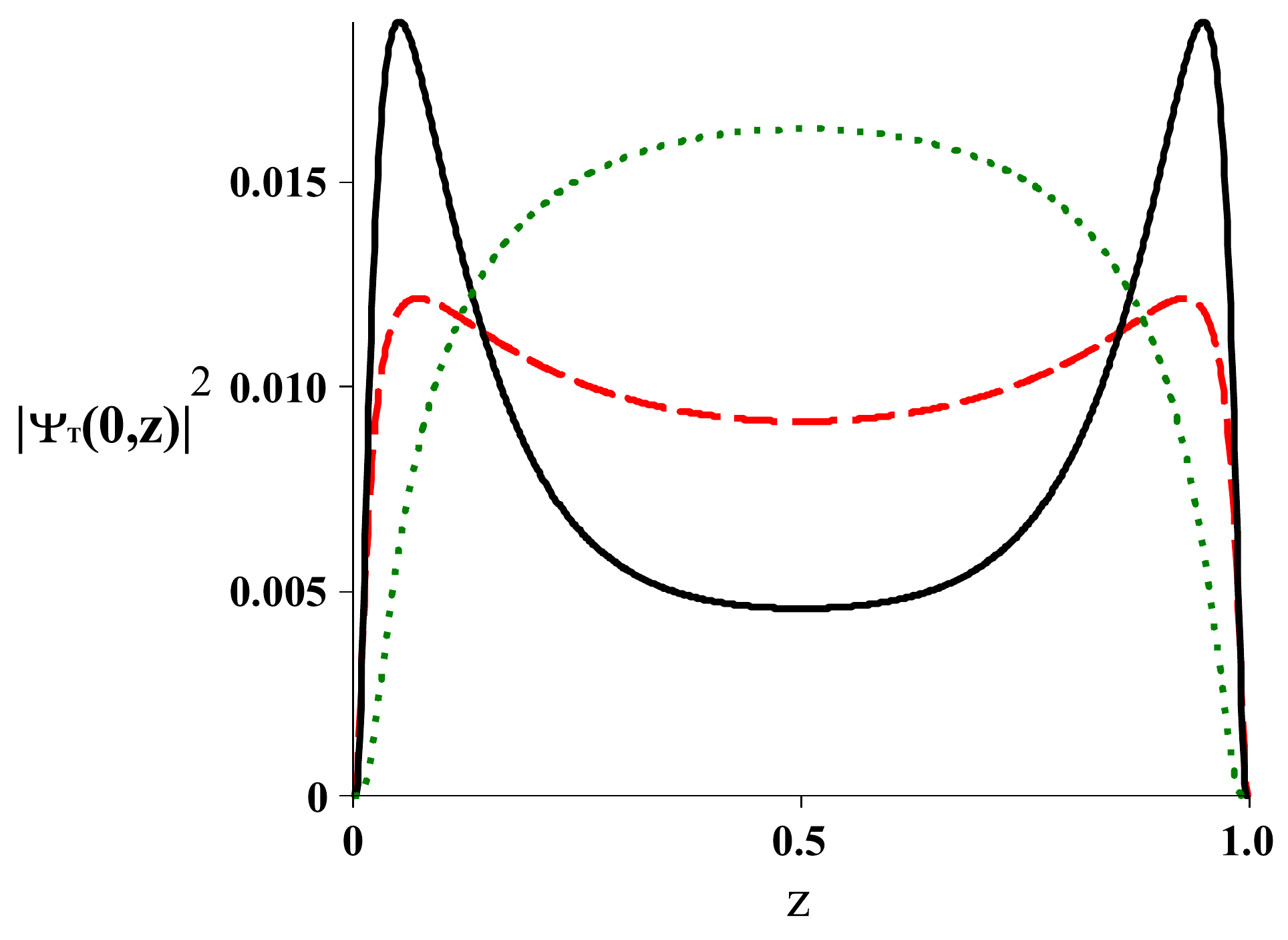}  
\caption{The longitudinal (left) and transverse (right) light-cone
  wavefunctions squared corresponding to the BG fits with (solid) and
  without (dashed) additional end-point
  enhancement in the transverse wavefunction. The BG parameterization
  of Ref.~\cite{Forshaw:2003ki} is also shown as the dotted curve.
All curves evaluated at $r=0$.}
\label{fig:LCWF_r0_FitZEUS}
\end{figure}

In Figure~\ref{fig:LCWF_FitZEUS}, we show the wavefunctions
corresponding to the improved fit. Compared to Figure~\ref{fig:BG_LCWF},
the end-point enhancement is quite distinctive. Figure
\ref{fig:LCWF_r0_FitZEUS} shows the wavefunctions for both fits (the
solid and dashed curves) at
$r=0$, plotted as a function of $z$. Note that they are almost indistinguishable
from
each other in the longitudinal case. For comparison, also shown on this figure
is the
result of the original BG parameterization \cite{Forshaw:2003ki}: it
gives the dotted curves.

The extent to which the data require any additional end-point
enhancement should be set in context since the $Q^2 \to 0$ limit suffers
from the greatest theoretical uncertainty. For example, we
characterize non-perturbative effects in the photon wavefunction
through a single parameter (the quark mass) and it is unclear how a
more sophisticated treatment would affect our conclusions. Moreover,
it is also possible to improve the quality of fit to the data without
appealing to Eq.~(\ref{EG}) by increasing the value of the diffractive
slope, $B$, at $Q^2=0$. Specifically, we acheived a $\chi^2/$degree of
freedom of $67/72$ after increasing $B$ by $15\%$ at $Q^2=0$ relative to the
value determined by Eq.~(\ref{Bslope}). 

We note that both our fits indicate that the data prefer a transverse
wavefunction
with enhanced contributions at $z \to 0,1$. This conclusion is valid regardless
of the
uncertainties on the forward slope parameter.

\begin{figure}
\centering
\includegraphics*[width=7.5cm]{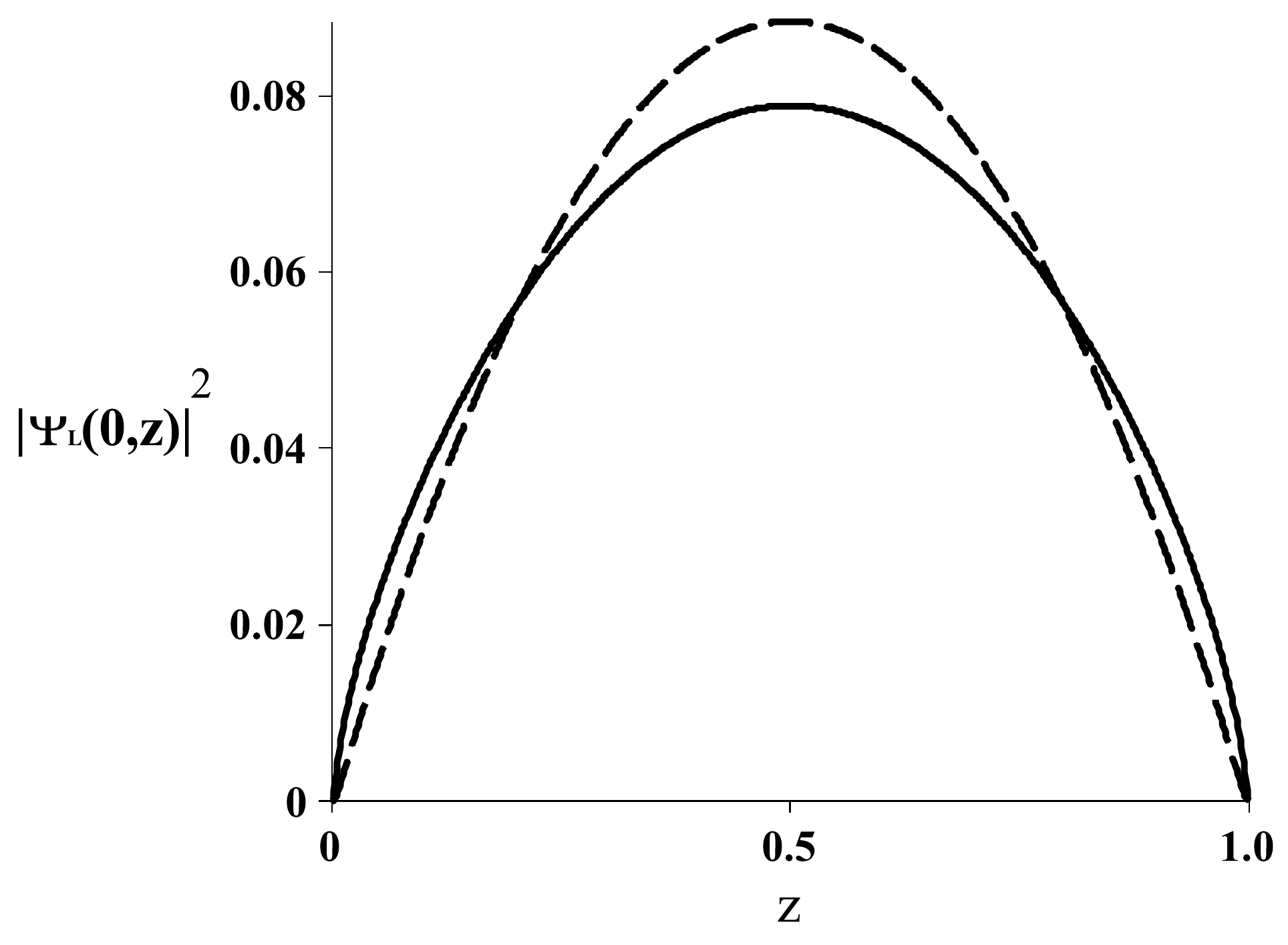}
\includegraphics*[width=7.5cm]{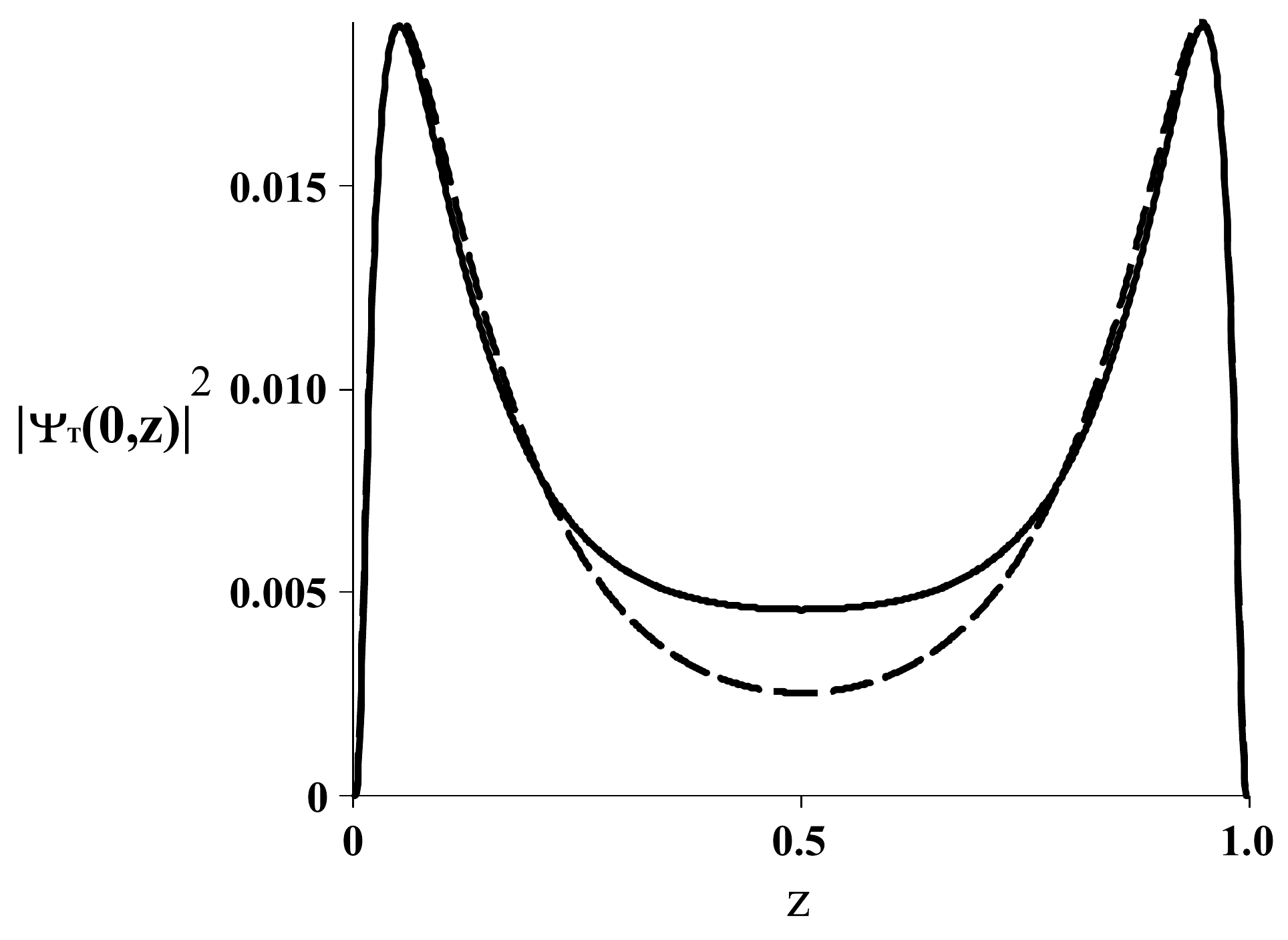}  
\caption{The longitudinal (left) and transverse (right) light-cone
  wavefunctions squared corresponding to the BG fits with additional end-point
  enhancement in the transverse wavefunction. The solid and dashed curves are
extracted after rescaling the data down by $5\%$ and $12\%$ respectively. 
Both curves evaluated at $r=0$.}
\label{fig:LCWF_r0_FitZEUS_rescale}
\end{figure}

Finally, we should remark that it is possible to lower the $\chi^2$ of the fit
still further if we are prepared to rescale the data downwards by more than
$5\%$. For example, rescaling down by $12\%$ leads to a $\chi^2$/degree of
freedom equal to $62/70$. The dashed curve in Figure
\ref{fig:LCWF_r0_FitZEUS_rescale} confirms that the corresponding wavefunction
is not very different from the one obtained by rescaling the data down by
$5\%$. 

\section{The leading-twist distribution amplitude}
We can use the longitudinal wavefunction from our fits to extract the
corresponding leading twist-2 distribution amplitude, $\varphi(z,\mu)$~\cite{Brodsky:1994kf}: 
\begin{equation}
\varphi(z,\mu) \sim \int \d^2\mathbf{k} \;
\Theta(|\mathbf{k}|<\mu) \tilde{\phi}_L(\mathbf{k},z)
\label{Leading-twist-DA-LCWF-k}
\end{equation}
where 
\begin{equation}
\tilde{\phi}_L(\mathbf{k},z) \sim \int \d^2\mathbf{r}  \; e^{-i
\mathbf{k}.\mathbf{r}} \phi_L (r,z)~,
\label{Fourier-transform-2d}
\end{equation}
which implies
\begin{equation}
\varphi(z,\mu) \sim \int \d r \; \mu J_1(\mu r) \phi_L(r,z) \;.
\label{Leading-twist-DA-non-asymp}
\end{equation}
Substituting for the scalar wavefunction from the previous section
gives
\begin{equation}
\varphi(z,\mu) \sim \left( 1 - \mathrm{e}^{-\mu^2/\Delta(z)^2}\right)
  \mathrm{e}^{-m_f^{2}/\Delta(z)^2} [z(1-z)]^{b_L}~,
\end{equation}
where $\Delta(z)^2 = 8[z(1-z)]^{b_L}/R_L^2$.
We note that our DA is very slowly varying with $\mu$ for $\mu >
1$~GeV. This means that our parameterization neglects the
perturbatively known
$\mu$-dependence of the DA and can thus be viewed as a
parameterization of the DA at some not too large a value of $\mu$. This
is reasonable given the limited $Q^2$ range of the HERA data to which
we fit (i.e. $\sqrt{Q^2} < 6$~GeV).

To compare with existing theoretical predictions for the DA, we
compute moments of our DA, i.e.
\begin{equation}
\langle \xi^n \rangle_{\mu} = \int_0^1 \d z \; \xi^n \varphi(z,\mu)~. 
\end{equation}
The $n=0$ moment is fixed by the decay constant constraint and is not a
prediction\footnote{It is equivalent to Eq.~(\ref{longdecay}) with the
  higher-twist $m_f^2-\nabla_r^2$ terms set to zero.}, we therefore follow convention and normalize the DA
according to
\begin{equation}
\int_0^1 \d z ~\varphi(z,\mu) = 1~.
\end{equation}
Our results are compared with the existing predictions in Table \ref{tab:moments-mu} and are in very good agreement with the
expectations based on QCD sum rules and from the lattice. Also shown
for comparison is the prediction based upon the old BG wavefunction
used in Ref.~\cite{Forshaw:2003ki}, which does not fit the HERA data. The
predictions in that case are rather similar to those of
Ref.~\cite{Choi:2007yu} using the light-front quark model.

\begin{figure}
\centering
\includegraphics*[width=10cm]{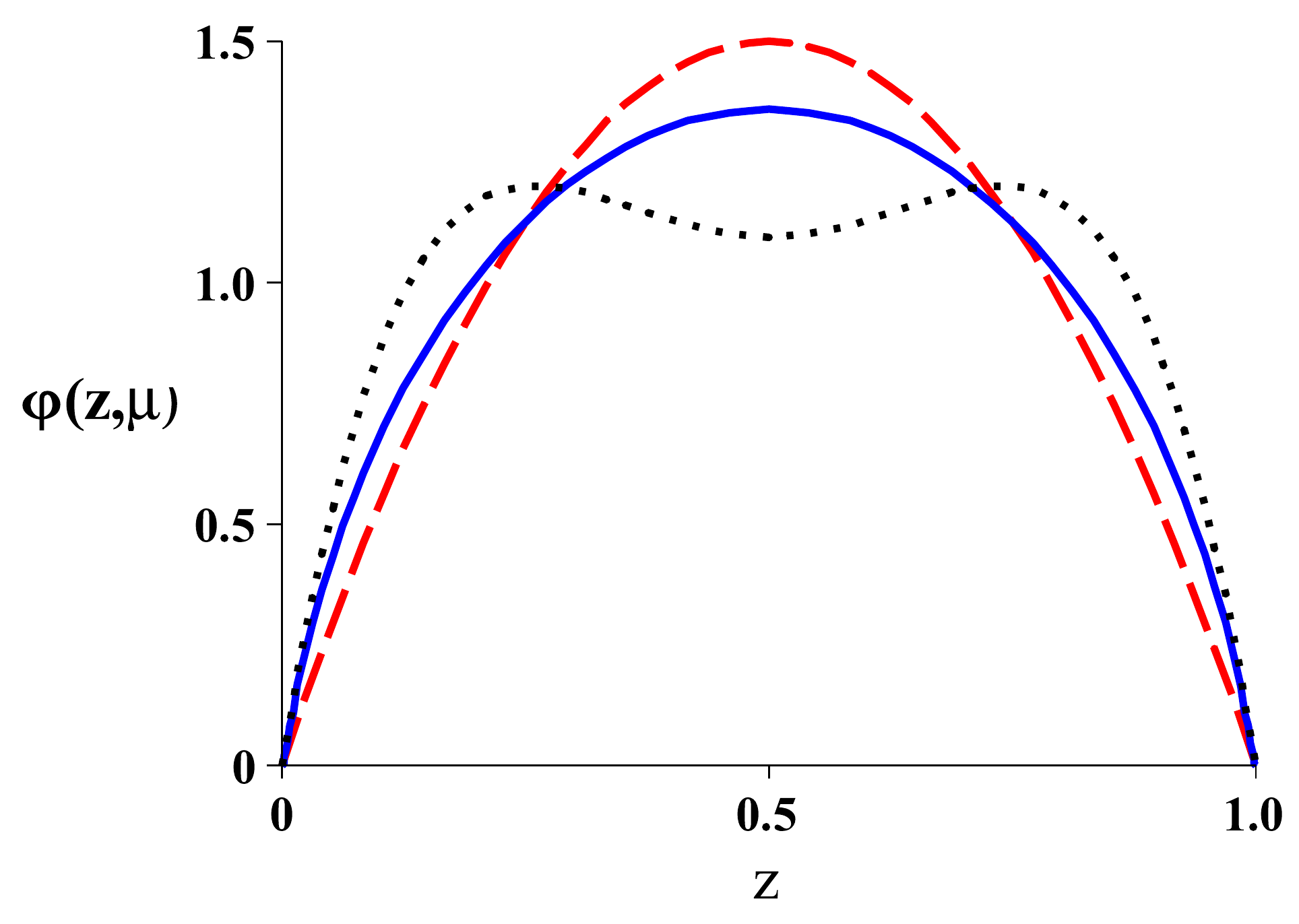}
\caption{The extracted DA at $\mu=1$ GeV (solid) compared to the DA at $1$ GeV  of Ref. \cite{Ball:1996tb} (dotted)
and the asymptotic DA (dashed).}
\label{fig:DA_1GeV}
\end{figure}

Finally, in Figure~\ref{fig:DA_1GeV} we compare our DA with that
predicted by Ball and Braun~\cite{Ball:1996tb}. The agreement
is reasonable given that in Ref.~\cite{Ball:1996tb}, the expansion in
Gegenbauer polynomials is truncated at low order, which is presumably
responsible for the local minimum at $z=1/2$. Certainly the two
distributions are broader than the asymptotic prediction $\sim 6z(1-z)$.

\begin{table}[h]
\begin{center}
\textbf{Moments of the leading twist DA at the scale $\mu$}
\[
\begin{array} 
[c]{|c|c|c|c|c|c|c|c|c}\hline
\mbox{Reference} & \mbox{Approach} & \mbox{Scale}~\mu &\langle \xi^2
\rangle_{\mu}&\langle \xi^4 \rangle_{\mu}&\langle \xi^6 \rangle_{\mu}&\langle
\xi^8 \rangle_{\mu}&\langle \xi^{10} \rangle_{\mu} \\ \hline
\mbox{(This paper)} & \mbox{Best fit}&\sim 1~\mbox{GeV} &0.227&
0.105&0.062&0.041&0.029\\ \hline
\mbox{(This paper)} & \mbox{BG fit}&\sim 1~\mbox{GeV} &0.229&
0.107&0.063&0.042&0.030\\ \hline
\mbox{(This paper)} & \mbox{Old BG prediction}&\sim 1~\mbox{GeV} &0.182&
0.072&0.037&0.022&0.014\\ \hline
\mbox{\cite{Choi:2007yu}}& \mbox{LFQM}&1~\mbox{GeV}&0.19 [0.21]& 0.08 [0.09] &
0.04 [0.05]& &\\
\hline
\mbox{\cite{Bakulev:1998pf}}&
\mbox{GenSR}&1~\mbox{GeV}&0.227(7)&0.095(5)&0.051(4)&0.030(2)&0.020(5) \\ \hline
\mbox{\cite{Chernyak:1983ej}}& \mbox{SR}&1~\mbox{GeV} &0.26&0.15 & & &  \\ \hline
\mbox{\cite{Ball:1996tb}} & \mbox{SR}&1~\mbox{GeV} &0.26(4)& & & &   \\ \hline
\mbox{\cite{Ball:2007zt}}&\mbox{SR}&1~\mbox{GeV} &0.254& & & &  \\ \hline
\mbox{\cite{Ball:2004ye}}&\mbox{SR}&1~\mbox{GeV} &0.23\pm^{0.03}_{0.02}&
0.11\pm_{0.02}^{0.03}& & &  \\ \hline
\mbox{\cite{Boyle:2008nj}}&\mbox{Lattice} &2~\mbox{GeV} &0.24(4)& & & &  \\
\hline 
 & 6z(1-z)
&\infty&0.2&0.086&0.048&0.030&0.021 \\ \hline
\end{array}
\]
\end{center}
\caption {Our extracted values for  $\langle \xi^n \rangle_{\mu}$, compared to
predictions based on the light-front quark model (LFQM), QCD sum rules (SR),
Generalised QCD Sum Rules (GenSR) or lattice QCD. Two predictions are given for
each moment in the LFQM approach; one corresponding to a harmonic oscillator
potential and the other (in square brackets) a linear potential.}
\label{tab:moments-mu}
\end{table}

\section{Conclusions}
The dipole model of diffractive photoproduction has been used
successfully to describe a large body of data \cite{Forshaw:2006np}. In this
paper we
have used it, together with accurate data on
$\rho$-meson photoproduction collected at the HERA collider
\cite{Chekanov:2007zr, Collaboration:2009xp}, in order to extract the
$\rho$ meson's light-cone wavefunction. The data require qualitatively
different behaviour for the two meson polarizations. In particular,
there is evidence for an enhancement of the end-point contributions to the
wavefunction for transversely polarized mesons. We extracted the leading-twist
$\rho$-meson distribution amplitude for longitudinal polarization and
found it to agree well with predictions based
on QCD sum rules and from the lattice.

\section{Acknowledgements}
We thank Patricia Ball, Aharon Levy, Paul Newman and Mike Seymour for helpful
discussions. R.S. acknowledges the hospitality of the Particle Physics
Group of the University of Manchester where parts of this work were
carried out. We also thank the UK's STFC for financial support.

\bibliographystyle{JHEP}
\bibliography{Fitrhov2}

\end{document}